\documentstyle[11pt,epsf]{article}
\newcommand{\beq}{\begin{equation}}
\newcommand{\eeq}{\end{equation}}
\newcommand{\beqa}{\begin{eqnarray}}
\newcommand{\eeqa}{\end{eqnarray}}

\newcommand{\dm}{\mbox{$\Delta{m}^{2}$~}}

\begin{document}
\thispagestyle{empty}

\renewcommand{\thefootnote}{\fnsymbol{footnote}}

\vskip 5pt

\begin{center}

{\large {\bf SOLAR NEUTRINO RATES, SPECTRUM, AND ITS MOMENTS : 
AN MSW ANALYSIS IN THE LIGHT OF SUPER-KAMIOKANDE RESULTS
}}

\vskip 10pt

{\sf Srubabati Goswami$^a$
\footnote{E-mail address: sruba@tnp.saha.ernet.in}},  
{\sf Debasish Majumdar$^{a,b}$
\footnote{E-mail address: debasish@tnp.saha.ernet.in}},  
and 
{\sf Amitava Raychaudhuri$^b$
\footnote{E-mail address: amitava@cubmb.ernet.in}}  

\vskip 8pt
$^a${\em Saha Institute of Nuclear Physics,\\ 
1/AF Bidhannagar, Calcutta 700 064, India. }\\

$^b${\em Department of Physics, University of Calcutta,\\ 
92 Acharya Prafulla Chandra Road, Calcutta 700 009, India. }

\vskip 20pt

{\bf ABSTRACT}

\end{center}


We re-examine  MSW solutions of the solar neutrino problem in a
two flavor scenario taking (a) the results on total rates and the
electron energy spectrum from the 1117-day SuperKamiokande (SK)
data and  (b) those on total rates from the Chlorine and Gallium
experiments. We find that the SMA solution gives the best fit to
the total rates data from the different experiments. One new
feature of our analysis is the use of the moments of the SK
electron spectrum in a $\chi^2$ analysis. The best-fit to the
moments is broadly in agreement with that obtained from a direct
fit to the
spectrum data and prefers a $\Delta m^2$ comparable to the SMA fit to
the rates but the required mixing angle is larger.  In the
combined rate and spectrum analysis, apart from varying the
normalization of the $^8$B flux as a free parameter and
determining its best-fit value we also obtain the best-fit
parameters when correlations between the rates and the spectrum
data are included and the normalization of the $^8$B flux held
fixed at its SSM value. We observe that the correlations between
the rates and spectrum data are important and the goodness of fit
worsens when these are included. In either case, the best-fit
lies in the LMA region.

\vskip 40pt

\begin{center} 
\end{center}

\newpage

\renewcommand{\thefootnote}{\arabic{footnote}}

\setcounter{footnote}{0}

\section{Introduction}

There is now strong evidence in support of an oscillatory
behavior of neutrinos. The results on atmospheric neutrinos from
SuperKamiokande \cite {ska} find a comprehensive explanation in
terms of oscillations, indicating a non-zero neutrino mass. This
result and the solar neutrino problem, which has been held as a
signal for neutrino oscillations for long,  have created
much excitement in the community.

In this paper we examine the latest solar neutrino data from
SuperKamiokande \cite{sk1117} along with those from the
radiochemical Chlorine and Gallium experiments assuming that MSW
resonant flavor conversion \cite{msw} is operative.  The above
experiments have presented the measured arrival rates
\cite{solar} which, in all the cases, are less than the
predictions from the Standard Solar Model (SSM).  In addition,
SuperKamiokande (SK) has provided the observed electron energy
distribution \cite{sk1117,sksolar}. We use these data to test the
consistency of the MSW mechanism  taken together with the SSM
predictions in a two flavor scenario.  There are several recent
MSW analyses of the SK solar neutrino data
\cite{sk825} - \cite{gg4}.  Although the data
fitting method followed in all of them is to minimize a $\chi^2$
function, the details of the statistical procedure used vary
among the different groups.  In this work, we indicate two
different ways of performing the statistical analysis for the
spectrum and the combined rate and spectrum data and check the
consistency of the best-fit values of the mass squared
differences and the mixing angles so obtained.  For the analysis
of the spectrum results we explore the possibility of using its
moments as variables for fitting the data.  The main advantage of
these moments is that they probe the shape of the spectrum in a
manner independent of the $^8$B flux normalization uncertainties.
For the combined analysis of the rates and the spectrum, apart
from the standard procedure of varying the $^8$B flux
normalization and treating these two sets of data as independent,
we also adopt a second method which takes into account the
correlations among the rates and the spectrum data.

In addition to the best-fit values of the oscillation parameters,
we present the 99\% C.L. and 90\% C.L. allowed regions
and the goodness of fit (g.o.f)
of a particular solution.  
By g.o.f. is meant the probability that the $\chi^2$
will exceed $\chi^2_{min}$. When presenting the allowed region we
take $\chi^2_{min}$ to be the value at the {\em global} minimum
in that region\footnote{The other approach is to present the
allowed regions with respect to the {\em local} minimum.}.  For the
neutrino fluxes and the neutrino production positions within the
sun we use the BP98 solar model \cite{bp98}. We consider
oscillation of $\nu_e$ to a sequential ($\nu_\mu$ or $\nu_\tau$)
neutrino.

This paper is organized as follows. In the next section we
present the formulae for oscillation of neutrinos with the
inclusion of matter effects both in the sun and during their
passage through the earth. In section 3 we use the data on the
total solar neutrino rates as measured at the Chlorine, Gallium,
and SuperKamiokande (1117-day data) detectors to obtain the
best-fit values of the neutrino mass splitting and the mixing
angles. In section 4 we consider the electron energy spectrum
observed at SuperKamiokande. Using the MSW predictions, we obtain
the best-fit values from a direct fit to the data as well as from
a fit to the normalized moments.  In section 5 we use both the
total rates data and the SK electron energy spectrum data to make
a combined fit. As noted earlier, here we allow the normalization
of the $^8$B spectrum to vary and compare the results with those
obtained when the SSM prediction for this normalization is used,
allowing the inclusion of correlations between the total
rates and the observed spectrum {\em via}  astrophysical
uncertainties. We end in section 7 with a summary, discussions,
and conclusions.

\section{Oscillation Probability}

In this work we restrict ourselves to the simplest case of mixing
between two neutrino flavors.  Assuming the neutrino mass
eigenstates, $\nu_{i}$, reaching the earth to be incoherent
\cite{dighe}, the survival probabaility of an electron neutrino
can be written as
\begin{equation}
P_{ee} = \sum_{i} P_{ei}^{S} P_{ie}^{E},
\label{pg}
\end{equation}  
where $P_{ei}^{S}$  is the probability of an electron neutrino
state to transform into the $i$-th mass state at the solar
surface and $P_{ie}^E$ is the conversion probability of the
$i$-th mass state to the $\nu_e$ state after traversing the
earth.  One can express $P_{ee}$ in terms of the day-time (i.e.
no earth matter effect) probability $P_{ee}^{D}$ as\footnote
{This expression is not applicable for maximal mixing
($\cos2\theta$ =0) \cite{guth} and for this case we use eq.
(\ref{pg}) directly.}
\begin{equation}
P_{ee} = P_{ee}^{D} + \frac{(2 P_{ee}^{D} - 1)(\sin^2 \theta -
P_{2e}^E)} {\cos{2\theta}},
\label{pee} 
\end{equation}
where
\begin{equation}
P_{ee}^{D} = 0.5+[0.5
-\Theta(E_\nu - E_{A})X]{\cos2\theta_{M}}{\cos2\theta}, 
\label{pf}
\end{equation}
$\Theta$ being the Heaviside function. $\theta$ is the mixing
angle in vacuum and $\theta_{M}$ is the mixing angle in matter
given by
\begin{equation}
\tan 2\theta_{M} =  \frac{\dm\sin 2\theta}{\dm\cos
2\theta - 2\sqrt{2}G_{F}n_{e}E_\nu}.
\label{thetam}
\end{equation}
Here $n_{e}$ is the ambient electron density, $E_\nu$ the
neutrino energy, and \dm (= ${m_{2}^2 - m_{1}^2}$) the mass squared
difference in vacuum.
\begin{equation}
E_{A} = {\Delta{m^{2}}\cos2\theta}/{2\sqrt{2}G_{F}n_{e}|_{pr}},
\end{equation}
gives the minimum $\nu_e$ energy that can encounter a resonance
inside the sun, $n_{e}|_{pr}$ being the electron density at the
point of production.  $X$ is the jump-probability between the
mass eigenstates and for an exponential density profile, as is
approximately the case in the sun, it is given by \cite{petcov}
\begin{equation}
X = \frac{ \exp~[- \pi \gamma_{R} (1 - \cos{2\theta})] - \exp~[-2 \pi
\gamma_{R}]} {1 - \exp[-2 \pi \gamma_{R}]}, 
\label{petcov}
\end{equation}
where $\gamma_{R} = \gamma \cos{2\theta}/\sin^22\theta$ and
\begin{equation} 
\gamma  =  \frac {\pi}{4} \frac {\Delta m^2}{E_\nu}
\frac {\sin^2 2 \theta }{\cos 2 \theta} \frac {1}{|\frac {d
\; \ln n_e} {d r}|_{res}}. 
\end{equation}
For calculating $P_{2e}^E$ in eq. (\ref{pee}) we treat the earth
as a slab of constant density (4.5 gm/cc). We have verified that
this is a reasonable approximation since the location of the SK
detector ensures that only in a rather small fraction of the time
does the neutrino pass through the denser core. 

The definition of $\chi ^2$ used by us in the following sections is,
\begin{equation}
\chi^2 =
\sum_{i,j} \left(F_i^{th} - F_i^{exp}\right)
(\sigma_{ij}^{-2}) \left(F_j^{th} - F_j^{exp}\right).
\label{chi}
\end{equation}
Here $F_{i}^{\xi}= {T_i^{\xi}}/{T_{i}^{BP98}}$ where $\xi$ is
$th$ (for the theoretical prediction with oscillations) or $exp$
(for the experimental value) and $T_i$ stand for the quantities
being fit (total rates from different experiments, electron
energy spectrum for different energy bins, etc.).  The error
matrix $\sigma_{ij}$ contains the experimental errors, the
theoretical errors and their correlations.

\section{Total rates}

In this section we perform an analysis of the total rates as
measured at the various experiments.  The data that we use 
for the total rates are given in Table 1.  In
particular, we use the 1117-day SK data.  For the Ga experiments
we take the weighted average of the SAGE and Gallex results.
Because SK has better statistics we do not include the Kamiokande
results.

\begin{table}[htb]
\begin{center}
\begin{tabular}{|c|c|c|c|}
\hline
Experiment & Chlorine & Gallium & SuperKamiokande \\ \hline
$\frac{\rm Observed \;\; Rate}{\rm BP98 \;\; Prediction}$ & 
$0.33 \pm 0.029$ & $0.562 \pm 0.043$ & $0.465 \pm 0.015$ \\ \hline
\end{tabular}
\end{center}

\begin{description}
\item{Table 1:} The ratio of the observed solar neutrino rates to
the corresponding BP98 SSM predictions used in this analysis. The
results are from Refs. \cite {solar} and \cite {sk1117}.  For
Gallium, the weighted average of the SAGE and Gallex results has
been used.
\end{description}
\end{table}

To get the best-fit values of the parameters, we minimise a
$\chi^2$ function defined as in eq. (\ref{chi}). For evaluating
the error matrix, $\sigma_{ij}$, we use the procedure described
in \cite{flap}. Our best-fit results for the total rates are
summarized in Table 2.  These fits have 1 degree of freedom (3
experimental data points -- 2 parameters). It is clear from this
Table that the small mixing angle solution for the sequential
neutrino case gives by far the best fit to the total rates data.

\begin{table}[htb]
\begin{center}
\begin{tabular}{|c|c|c|c|c|c|}
\hline
No. of days of&Nature of & $\Delta m^2$ &
$\sin^2\theta$&$\chi^2_{min}$& Goodness\\
SK running&Solution & in eV$^2$&  & & of fit\\
\hline
&SMA & $5.19 \times 10^{-6}$&$1.53 \times 10^{-3}$ & 
0.30 & 58.38\%  \\ \cline{2-6}
1117&LMA & $1.42 \times 10^{-5}$ & 0.22 & 
3.52 & 6.06\% \\  \cline{2-6}
&LOW & $9.97 \times 10^{-8}$ & 0.38 & 6.64 & .997\% 
 \\ \hline
&SMA & $5.33 \times 10^{-6}$&$1.50 \times 10^{-3}$ & 
0.06 & 80.64\%  \\ \cline{2-6}
825&LMA & $1.40 \times 10^{-5}$ & 0.23 & 
4.31 & 3.78\%  \\  \cline{2-6}
&LOW & $9.98 \times 10^{-8}$&$0.38$ & 
7.39 & 0.65\%  \\ \hline
\end{tabular}
\end{center}

\begin{description}
\item{Table 2:} The best-fit values of the parameters,
$\chi^2_{min}$, and the g.o.f. for fits to the total
rates of the different experiments. SMA, LMA, LOW stand for the
Small Mixing Angle, Large Mixing Angle and Low mass-low $\chi^2$ 
regions, respectively. The results for the earlier 825-day SK data
(0.475 $\pm$ 0.015) are also presented for comparison.
\end{description}
\end{table}

It is seen from Table 2 that the best-fit values obtained for the
1117-day data are not markedly different from those obtained for
the 825-day data.  This indicates that the best-fit points
obtained from the analysis of total rates are quite robust and
are not expected to change drastically with more data from SK.
However, the quality of fit is quite sensitive to these small
changes: the g.o.f. has become a little poorer for the SMA solution
while for both the LMA and LOW solutions it has improved with the
accumulation of more SK data.

In Fig. 1 we show the 99\% C.L. and 90\% C.L. contours in the
$\sin^2\theta - \Delta m^2$ plane for the MSW solution.  There is
no solution in the {\em dark side} ($\sin^2\theta > 0.5$) from
the analysis of total rates. Note that in \cite{murayama} allowed
regions were obtained in the dark side from the total rates
analysis while in \cite{ggds} no allowed region was found for
$\theta > \pi/4$.  We agree with \cite{ggds}; a possible origin
of this is that the same numerical density profile of the sun
from BP98 has been used in these analyses.  Our results for the
total rates for the 1117 day data are in agreement with the
analysis given in
\cite{gg2000}. 

\section{Observed spectrum and its moments}

In addition to the total rates, SK has provided the number of
events (normalized to the SSM prediction) in 18 electron recoil
energy bins of width 0.5 MeV in the range 5.0 MeV to 14 MeV and a
19th bin which covers the events in the range 14 to 20 MeV
\cite{sk1117}. The systematics of the first bin are still under
study and for our analysis we do not use it.

\subsection{Observed spectrum}

In this subsection we present the results obtained by directly
fitting the SK spectral data. The theoretical predictions are
calculated bin by bin and in the fitting procedure, in addition
to the neutrino mixing parameters $\Delta m^2$ and $\sin^2\theta$, we
also allow the absolute normalization of the $^8$B flux, $X_B$,
to vary\footnote{$X_B$ = 1, for the SSM.}.  The error matrix
$\sigma_{ij}$ used by us (see eq. (\ref{chi})) is \cite{valle}
\begin{equation}
(\sigma_{ij}^2)_{sp} = \delta_{ij}(\sigma^2_{i,stat} +
\sigma^2_{i,uncorr}) + \sigma_{i ,exp} \sigma_{j,exp} +
\sigma_{i,cal} \sigma_{j,cal},
\end{equation}
where we have included the  statistical error, the uncorrelated
systematic errors and the energy-bin-correlated experimental
errors \cite{skspec} as well as those from the calculation of the
shape of the expected spectrum \cite{shape}. Since we vary the
normalization of the $^8$B flux we do not include its 
astrophysical uncertainties separately.

The best-fit point from this analysis is found to be
\begin{itemize}
\item $\Delta m^2 = 2.29 \times 10^{-6}$ eV$^2$, $\sin^2\theta =
0.009$, $X_{B} = 1.4$, $\chi^2_{min} = 9.46 $, g.o.f. = 85.23\%.
\end{itemize}

For these values of $\Delta m^2$ and $\sin^2\theta$, choosing
$X_B = 1$, the data on total rates give a bad fit ($\chi^2$ =
73.82) as the high energy $^8$B neutrinos get suppressed more
than observed. For the spectrum fit this problem can be avoided
by a high $X_B$ $>$ 1.  Since the $^8$B flux normalization is
allowed to vary, a large range of $\Delta m^2$ and $\sin^2\theta$
including the $\theta > \pi/4$ region remains allowed by the
spectral data as is shown in Fig. 2. Only the area inside the
contours is {\em disallowed} at 90\% C.L. from spectral data
analysis.

To further examine this mismatch between the fits to the total
rates and those to the SK spectral data, in Fig. 3 we present the
behavior of $\chi^2$ as obtained from the spectral analysis if we
keep the parameters $\sin^2\theta$ and $\Delta m^2$ in the SMA,
LMA, and LOW regions of the fit to the total rates.  In Figs.
3(a), 3(b), and 3(c) are shown the variation of the $\chi^2$ from
the spectral data with $\sin^2\theta$ lying in  the SMA, LMA, and
LOW regions respectively. For any chosen $\sin^2\theta$, we let
$\Delta m^2$ vary over the corresponding range permitted by the
fit to the total rates at 99\% C.L. (from Fig. 1) and plot the
minimum value found.  We consider two cases,
\begin{itemize} 
\item the $^8$B normalization
is held fixed at its SSM value (solid curves) 
\item the $^8$B normalization is 
permitted to vary (broken curves).
\end{itemize}
For Figs. 3(d), 3(e), and 3(f),
the roles of $\sin^2\theta$ and $\Delta m^2$ are interchanged.
Fig. 3 indicates that if we allow the $^8$B normalization to vary
then the $\Delta m^2$ and $\sin^2\theta$ allowed at 99\% C.L.
from the total rates are allowed at 90\% C.L. from the spectral
analysis. 

Figs. 2 and 3 lead us to the conclusion that, at the present
moment, the SK electron spectral data do not provide tight
controls over the allowed parameter range. 

\subsection{Moments of spectrum}

Since the absolute normalization of the $^8$B flux is not
precisely known, it is of interest to look for variables which
probe neutrino oscillation effects in the data in a manner immune
to this uncertainty.  Normalized moments of the observed electron
spectrum can be useful as one such set of variables
\cite{mom,othmom}. In practice, to compare with the data, it is
convenient to standardize with respect to the SSM predictions by
using 
\begin{equation}
M_n = \frac{\sum_i \left[\frac{N(E_i)}{\{N(E_i)\}_{SSM}}\right]
E_i^n}{\sum_i \left[\frac{N(E_i)}{\{N(E_i)\}_{SSM}}\right]},
\label{Mnssm}
\end{equation}
where $E_i$ is the mean energy of the $i$-th bin and  $N(E_i)$ is
the number of events in this bin. Depending on whether the
experimental or the theoretically predicted value of the variable
is under consideration, $N(E_i)$ is obtained either from
experiments or from the theoretical model under test.  It is
clear that these variables carry information about the shape of
the neutrino spectrum which, if oscillations are operative,
undergoes modification from the SSM prediction  due to the energy
dependence of the survival probability. It is obvious that the
above moments are independent of the absolute normalization of
the $^8$B flux.

We have calculated the moments of the 1117-day data on the
electron energy spectrum presented by SuperKamiokande
\cite{sk1117}.  These are presented in Table 3.  The error in the
higher moments increases rapidly with the order and the ones
beyond the sixth are not of much use.\\

\begin{table}[htb]
\begin{center}
\begin{tabular}{|c|c|c|}
\hline
Order of & Value of &  Calculated Error \\
Moment & Moment  &    in  Moment  \\
\hline
1& 10.25  &  0.33   \\
2 & 113.84 &  7.12   \\
3 & 1356.59 & 164.34  \\
4 & 17167.7 & 3616.65 \\
5 & 228876.2 & 75981.0 \\
6 & 3188064.0 & 1541763.0 \\
\hline
\end{tabular}
\end{center}

\begin{description}
\item{Table 3:} Moments of the observed electron energy spectrum
and their calculated errors obtained from the 
SK (1117 days) data.
\end{description}
\end{table}

Using these variables in a $\chi^2$-analysis we find that fitting
the first four moments results in the same best-fit values of $\Delta
m^2$ and $\sin^2\theta$ as those obtained using the first
five or six moments. The best-fit region\footnote{The $\chi^2$
remains unchanged when $\sin^2\theta$ is varied over the
indicated range} is found to be
\begin{itemize}
\item $\Delta m^2 = 6.94 \times 10^{-6}$ eV$^2$, $\sin^2\theta =
0.007 - 0.010$, $\chi^2_{min} = 0.0001 $, g.o.f = 99.995\%.
\end{itemize}

The very small value of $\chi^2_{min}$ for this fit should not
be regarded as a major success of the theory but rather reflects
the large errors associated with the moments as obtained from the
present data. It is gratifying that the best-fit values obtained
by this method are in the same broad region as those from fitting
the recoil electron energy spectrum.\\

\section{Combined fits to rates and spectrum}

In this section we present the results of the combined fit to the
total rates and the spectrum data. We have performed this global
fit by the following two methods:
\begin{itemize}
\item[(a)] We treat the rates and the electron
spectrum data as independent. In this approach we vary the $^8$B flux 
normalization as a free parameter.  
\item[(b)] We fix the $^8$B flux normalization at the SSM value
(=1) and include the correlations of the $^8$B flux uncertainty
between the rates and spectrum data. To our knowledge, this
approach has not been pursued in any previous analysis.
\end{itemize}

\subsection{Fits using the $^8$B flux normalization as a free
parameter}

For this case the definition of $\chi
^2$ is,
\begin{eqnarray}
\chi^2 &=&
\sum_{i,j=1,3} \left(F_i^{th} -
F_i^{exp}\right)
(\sigma_{ij}^{-2}) \left(F_j^{th} - F_j^{exp}\right) \nonumber \\
&& + \sum_{i,j=1,18} \left(X_B R_i^{th} -
R_i^{exp}\right)
(\sigma_{ij}^{-2})_{sp} \left(X_B R_j^{th} - R_j^{exp}\right),
\label{rspmd}
\end{eqnarray}
where the first term on the r.h.s is from the fit to the total
rates and the second from that to the spectral data.  As we allow
the normalization of the $^8$B flux to vary as a free parameter
we switch off the SSM astrophysical uncertainties arising because of
this component. Since it is the $^8$B flux that enters the
rates as well as the spectrum data, in this manner of fitting the
data the correlations between the rates and the spectrum are
absent; the error matrix is block diagonal and one can treat 
$\chi^2_{rate}$ and $\chi^2_{spectrum}$ as independent. There are
18 (= 21 -- 3) degrees of freedom in this case.  The best-fit
values we obtain are presented in Table 4.

\begin{table}[htb]
\begin{center}
\begin{tabular}{|c|c|c|c|c|c|}
\hline
Nature of & $\Delta m^2$ &
$\sin^2\theta$&$X_B$&$\chi^2_{min}$& Goodness\\
Solution & in eV$^2$&  & & & of fit\\
\hline
SMA & $4.81 \times 10^{-6}$&$5.92 \times 10^{-4}$ & 0.61 &
12.62 & 81.36\%  \\ \hline
LMA & $1.82 \times 10^{-5}$ & 0.18 & 1.39 &
10.97 & 89.56\%  \\ \hline
LOW & $1.0 \times 10^{-7}$&$0.38$ & 0.95 &
17.21 &50.87\%  \\ \hline
\end{tabular}
\end{center}
\begin{description}
\item{Table 4:} The best-fit values of the parameters, $\chi^2_{min}$,
and the g.o.f. for fits to the total rates measured at the Cl,
Ga, and SK detectors along with the electron energy spectrum from SK
(1117 days) when the $^8$B flux normalization factor $X_B$ is
allowed to vary. 
\end{description}
\end{table}

In Fig. 4a we show the 99\% and 90\% C.L. allowed regions for the
combined analysis of total rates and the observed electron spectrum. 
The best-fit points in these plots are obtained by varying $X_B$ in
addition to $\Delta m^2$ and $\sin^2\theta$.

\subsection{Fits including correlations between rates and
spectrum {\em via} $^8$B flux}
 
For this case we include the correlations between the theory errors
in the rate and the spectrum data.  This comes through the
$^8$B flux, as it enters both.  Since we include the SSM
astrophysical uncertainties in the $^8$B flux the normalization
factor for it is held fixed at the SSM value.  Now the individual
$\chi^2$ due to the spectrum and the rates cannot be summed
independently and the combined $\chi^2$ is defined as,
\begin{equation} 
\chi^2 =
\sum_{i,j=1,21} \left(F_i^{th} -
F_i^{exp}\right)
(\sigma_{ij}^{-2}) \left(F_j^{th} - F_j^{exp}\right),
\label{rtspmfit}
\end{equation}
where the $\sigma_{ij}$ is now a 21 $\times $ 21 matrix defined
in the following way,
\begin{itemize}
\item For  
$i, j$ = 1 \ldots 3 
\begin{equation}
\sigma_{ij}^2 =(\sigma^2_{ij})_{th} + (\sigma^2_{ij})_{exp},
\end{equation}
where
\begin{equation}
(\sigma^2_{i j})_{th} = \delta_{i j}\sum_{\alpha=1}^8 R^2_{\alpha i
}\,(\Delta C_{\alpha i })^2 + \sum_{\alpha, \beta =1}^8 R_{\alpha i}\,
R_{\beta j}
\sum_{k=1}^{e11} a_{\alpha k}\,a_{\beta k}\,(\Delta \ln X_k)^2\ .
\label{rterror}
\end{equation}
where the first term is due to the 
cross-section uncertainties and the second term ($\sigma_{ap}$) is
due to the astrophysical
uncertainties \cite{flap}.   
The off-diagonal elements in the error matrix come through $\sigma_{ap}$. 
$R_{\alpha i}$ denotes the contribution of the $\alpha$-th source
to the rate of the $i$-th experiment.  $a_{\alpha k} = {\delta
\ln \phi_{\alpha}}/{\delta \ln X_k}$, where $\delta \ln
\phi_{\alpha}$ is the error in the $\alpha$-th component of the
spectrum due to the input parameter $X_k$ \cite{bp98}.

\item For $i$ = 4 \ldots 21 and $j$ = 1 \ldots 3 
\begin{equation}
\sigma^2_{ij} = \sum_{\alpha=1}^{8} R_{ ^{8}{B} i} R_{\alpha j}
\sum_{k=1}^{11} a_{{^{8}{B}} k} a_{\alpha k} (\Delta \ln X_k)^2\ .
\end{equation}
 
\item For $i$ = 1 \ldots 3 and $j$ = 4 \ldots 21 
\begin{equation}
\sigma^2_{ij} = \sum_{\alpha=1}^{8} R_{{\alpha} i} R_{^{8}{B} j}
\sum_{k=1}^{11} a_{\alpha k} a_{^{8}{B} k} (\Delta \ln X_k)^2\ .
\end{equation}

\item for $i$ = 4 \ldots 21 and $j$ = 4 \ldots 21 
\begin{equation}
\sigma^2_{ij} = (\sigma_{ij}^2)_{sp} + R_{{^{8}{B}} i} R_{^{8}{B} j}
\sum_{k=1}^{11} a_{^{8}{B} k} a_{^{8}{B} k} (\Delta \ln X_k)^2\ .
\end{equation}

\end{itemize}
In this case the number of degrees of freedom is 19 (= 21 -- 2).
The $\chi^2_{min}$ and the best-fit values we obtain
are shown in Table 5.

\begin{table}[htb]
\begin{center}
\begin{tabular}{|c|c|c|c|c|}
\hline
Nature of & $\Delta m^2$ &
$\sin^2\theta$&$\chi^2_{min}$& Goodness\\
Solution & in eV$^2$&  & & of fit\\
\hline
SMA & $5.15 \times 10^{-6}$&$4.8 \times 10^{-4}$ & 
15.72 &67.58\%  \\ \hline
LMA & $2.23 \times 10^{-5}$ & 0.25 & 
14.53 & 75.19\%  \\ \hline
LOW & $1.00 \times 10^{-7}$&$0.38$ & 
17.62 & 54.79\%  
\\ \hline
\end{tabular}
\end{center}

\begin{description}
\item{Table 5:} The best-fit values of the parameters, $\chi^2_{min}$,
and the g.o.f. for fits to the total rates measured at the Cl,
Ga, and SK detectors along with the electron energy spectrum from SK
(1117 days). In this case the $^8$B flux is chosen as in the SSM
and the correlation between the rates and the spectrum due to
astrophysical uncertainties is included. 
\end{description}
\end{table}

We find that the fits are of poorer quality in this case as
compared to the previous one.  

In Fig. 4b we show the 99\% and 90\% C.L.  allowed regions for the
combined analysis of total rates and the observed electron
spectrum for MSW conversion to sequential neutrinos
including the correlations between the rates and the spectrum due
to the astrophysical uncertainties of the $^8$B flux normalization.

\section{Summary, Discussions, and Conclusions}

In this paper we have performed a detailed $\chi^2$-analysis of
the latest SK solar neutrino data together with the results from
the Cl and Ga experiments in terms of two-generation MSW
conversions of $\nu_e$ to sequential ($\nu_\mu$, $\nu_\tau$) 
neutrinos. 

Compared to the recent analyses in the literature 
\cite{sk825}-\cite{gg4}
there are two new features in our analysis. 
\begin{itemize}
\item  We fit
the observed electron energy spectrum data in two different ways,
exploring for the first time, the use of moments of the energy
spectrum in a $\chi^2$-analysis.  

\item The combined fits to the total
rates and spectral data are also performed in two different
manners. In the first, the $^8$B flux normalization is used as a
free parameter while in the other the SSM normalization is chosen
for it and correlations between the rates and spectral data due
to astrophysical uncertainties of the $^8$B flux are included.

\end{itemize}

We find that the two-generation MSW scenario can well explain the
data on total rates.  The solution in the SMA (Small Mixing
Angle) region is
preferred over the other possibilities although the quality of
the fit is poorer as compared to the one obtained using the
825-day SK data.

The best-fit from the spectrum data comes in a region disallowed
from the total rates. In this region the $^8$B neutrinos are
suppressed much more than required by the rates data.
For the analysis of the spectrum, the absolute
normalization of the $^8$B flux, $X_B$, has been permitted to be
greater than unity, thus effectively compensating the shortfall.
We have explored the use of  normalized moments of the observed
electron energy spectrum to signal MSW resonant flavour
conversion. These variables are independent of the absolute
normalization of the $^8$B flux and probe the effect of
oscillations on the spectral shape. This procedure is somewhat
handicapped by the large errors on the moments calculated from
the present data. However, the best-fit values obtained by the
two methods are more or less in agreement.

Similarly, for the two methods followed in the combined $\chi^2$
analysis of the rates and time averaged spectrum data, the best-fit values are
not much different.  The first approach gives a better fit
because we utilise the freedom of varying the $^8$B flux
normalization.  We remark that in the combined analysis, where
the $^8$B normalization is held fixed at the SSM value, the
correlations between the rates and the spectrum data are found to
be important and thus one should use caution regarding results
obtained  treating these as independent.
For both methods, the best-fit from the combined analysis falls in
the LMA region. Compared to the rates analysis the goodness of fit 
of the LOW(SMA) region increases(decreases). With the inclusion
of the day-night dependence of the data the goodness of fit in the
SMA region worsens further \cite{sk1117}.  

In this work we have not included the new GNO result \cite{gno} which 
is consistent with the Gallex and SAGE data. Thus its 
inclusion is not expected to affect  
the conclusions drastically. 
For illustration we give below the results   
of the global analysis of rates and spectrum including the GNO data.
We take the weighted average of Gallex and GNO and treat
SAGE as a separate experiment.
The best-fit values and $\chi^2_{min}$ that we get are:
\begin{itemize} 
\item $\sin^2 \theta = 5.26 \times 10^{-4},~~
\Delta m^2 = 5.28 \times 10^{-6}$ eV$^2,~~
X_{B} = 0.61,~~
\chi^2_{min} = 12.73$,~~
g.o.f = 85.21\%
~~~~(SMA)
\item $\sin^2 \theta$ = 0.18,~~
$\Delta m^2 = 2.48 \times 10^{-5}$ eV$^2$,~~
$X_{B}$ = 1.39,~~
$\chi^2_{min}$ = 11.55,~~
g.o.f = 90.39\%~~~~(LMA)
\item $\sin^2 \theta$ = 0.41,~~
$\Delta m^2 = 9.39 \times 10^{-8}$ eV$^2$,~~
$X_{B}$ = 0.89,~~
$\chi^2_{min}$ = 19.85,~~
g.o.f = 40.34\%~~~~(LOW)
\end{itemize}
Thus the global best-fit continues to be in the LMA region.  

In conclusion, we have probed the most recent solar neutrino data
on total rates and the observed electron energy spectrum  at SK
from various angles within the framework of MSW flavour
conversion. We find good fits in some situations but a degree of
uncertainty still remains since different fits do not prefer the
same values of the parameters. More data from the running and new
experiments, it is hoped, will further sharpen the results  in
the near future.\\

\parindent 0pt

{\large{\bf {Acknowledgements}}}\\

D.M. and A.R. are partially supported by the Eastern Centre for
Research in Astrophysics, India. A.R. also acknowledges a
research grant from the Council of Scientific and Industrial
Research, India. We would like to thank Sandhya Choubey for
pointing out an error in one of our computer codes and J.W.F.
Valle for drawing our attention to the updated analysis in
\cite{valle}.  S.G. would like to thank Plamen Krastev for many
helpful correspondences.

\newpage

\begin{center}
{\bf{\Large Figure Captions}}
\end{center}

Fig 1. The 99\% and 90\% C.L. allowed regions in the $\Delta m^2$
- $\sin^2\theta$ plane from the analysis of total rates for the
Chlorine and Gallium detectors and the 1117-day data from SK.
The best-fit points are also indicated. The dark side ($\sin^2 \theta 
> 0.5$) is indicated by the dashed line. 
\\

Fig 2. The 90\% C.L. allowed region in the $\Delta m^2$ - $\sin^2
\theta$ plane from the 1117-day SK recoil electron spectrum data.
The regions enclosed by the contours are {\em disallowed}. The
best-fit point is indicated. The dark side is to the right of the 
dashed line. \\

Fig. 3. The minimum $\chi^2$ for fits to the SK 1117-day recoil
electron spectrum as a function of $\sin^2\theta$ ($\Delta m^2$)
are shown in (a), (b), and (c) ((d), (e), and (f)) when the
parameter ranges are determined by the 99\% C.L. allowed regions
in the SMA, LMA, and LOW fits respectively to the total rates
data.  The solid (broken) curves are obtained when $X_B$ is held
fixed at its SSM value (allowed to vary). 
The dash-dotted line indicates the 90\% C.L. limit for 3 
parameters. See text for more
details.\\

Fig. 4. The 99\% and 90\% C.L. allowed region in the $\Delta m^2$
- $\sin^2 \theta$ plane from an analysis of the total rates from
the Chlorine and Gallium detectors and the 1117-day SK data taken
together with the 1117-day SK recoil electron spectrum.  The
normalization of the $^8$B flux is chosen as a free parameter in
(a) and held fixed at the SSM value in (b). In (b) the
correlations between the rates and spectrum data are included.
The best-fit points are also indicated. The dark side corresponds to the 
right of the dashed line.   
\begin{figure}[p]
\epsfxsize 15 cm
\epsfysize 15 cm
\epsfbox[25 151 585 704]{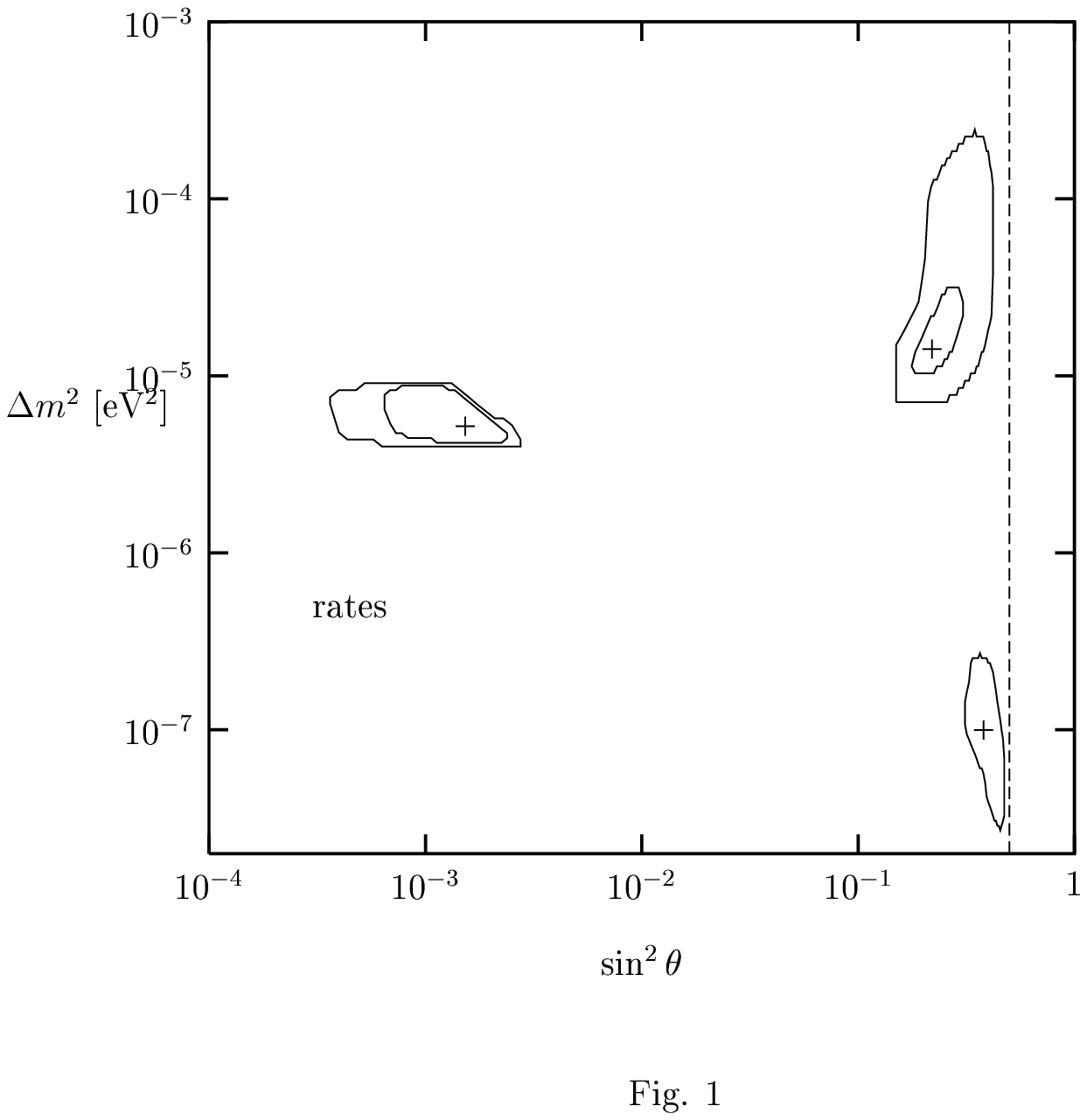}
\end{figure}

\begin{figure}[p]
\epsfxsize 15 cm
\epsfysize 15 cm
\epsfbox[25 151 585 704]{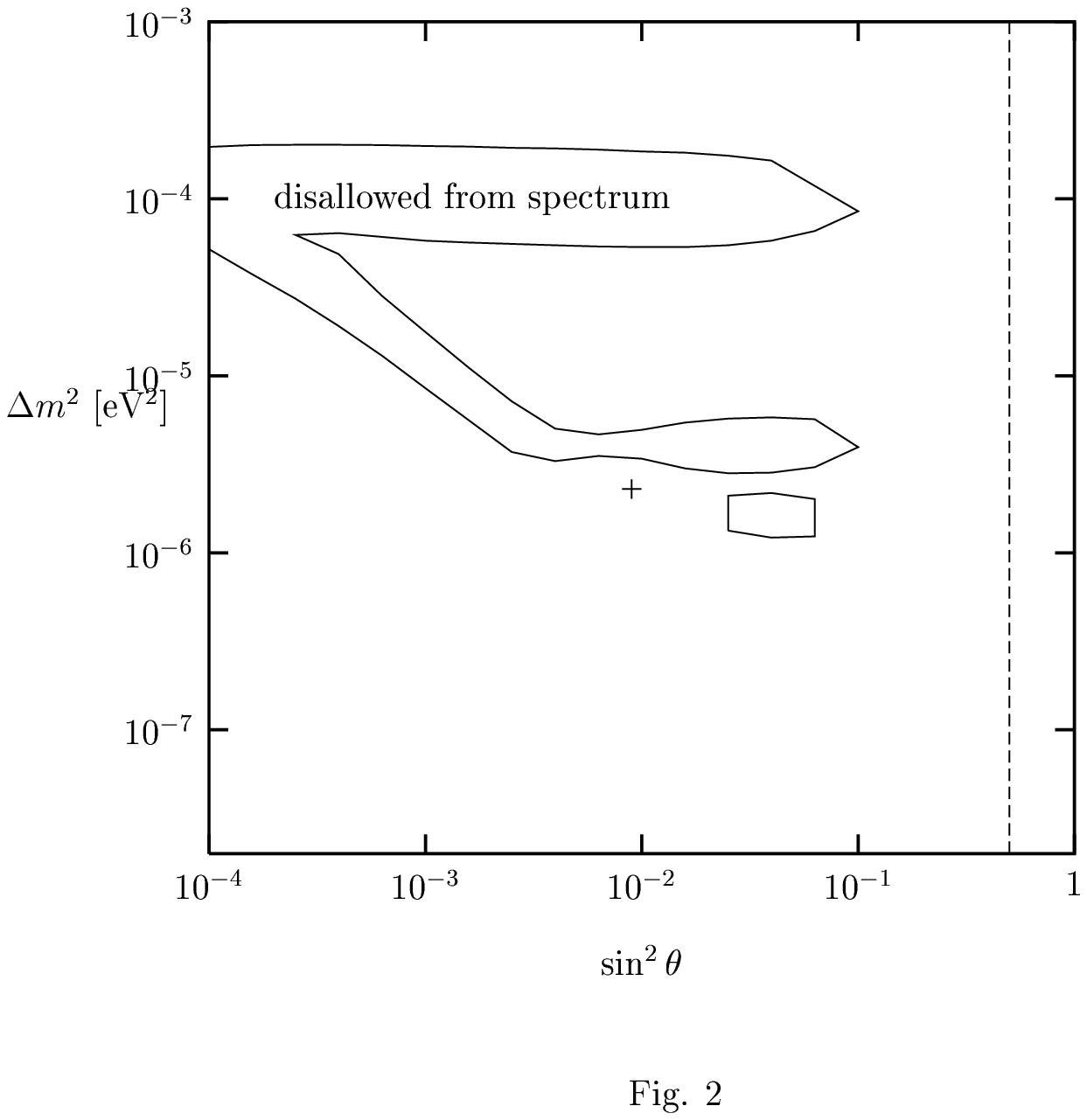}
\end{figure}

\begin{figure}[p]
\epsfxsize 15 cm
\epsfysize 15 cm
\epsfbox[25 151 585 704]{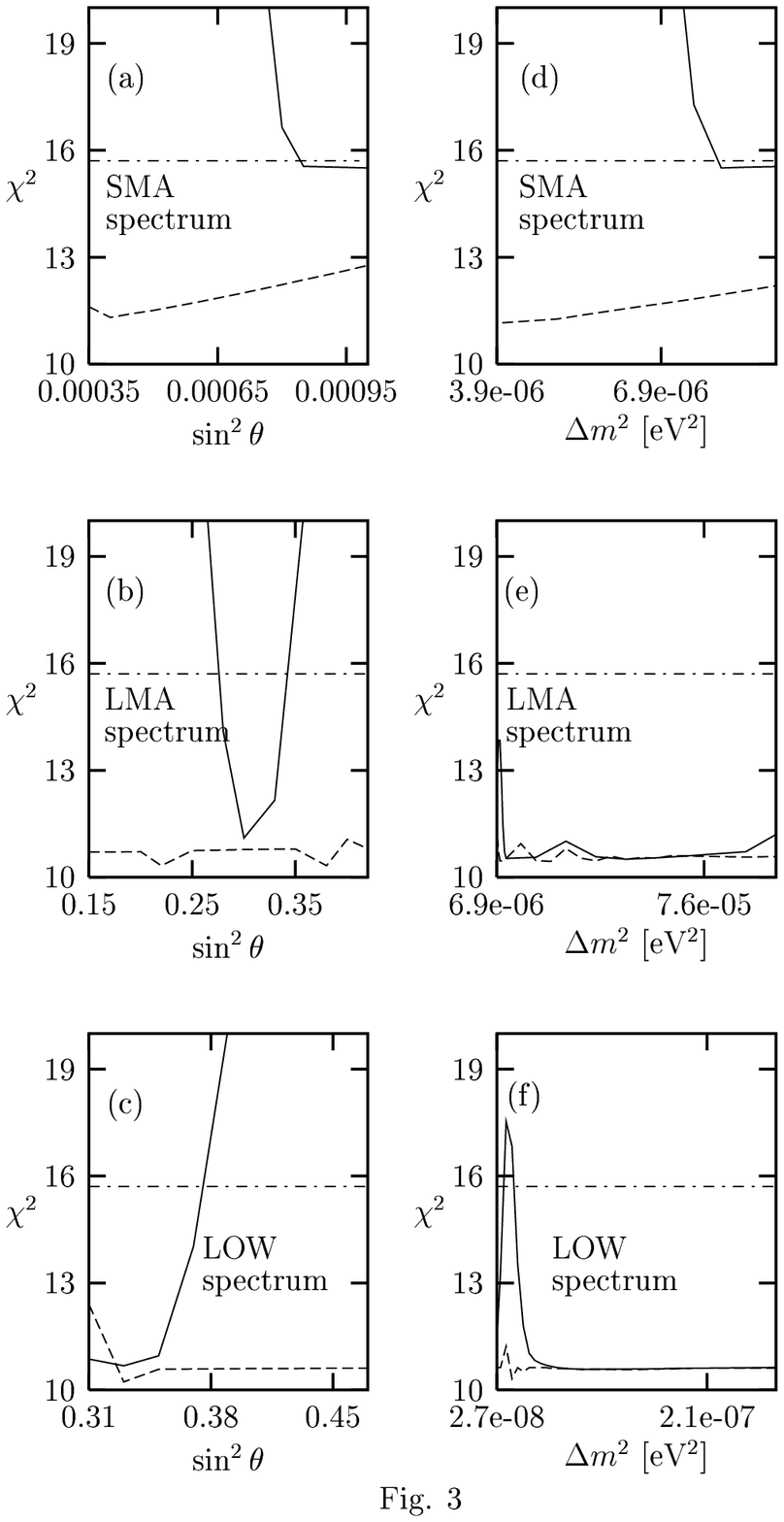}
\end{figure}

\begin{figure}[p]
\epsfxsize 15 cm
\epsfysize 15 cm
\epsfbox[25 151 585 704]{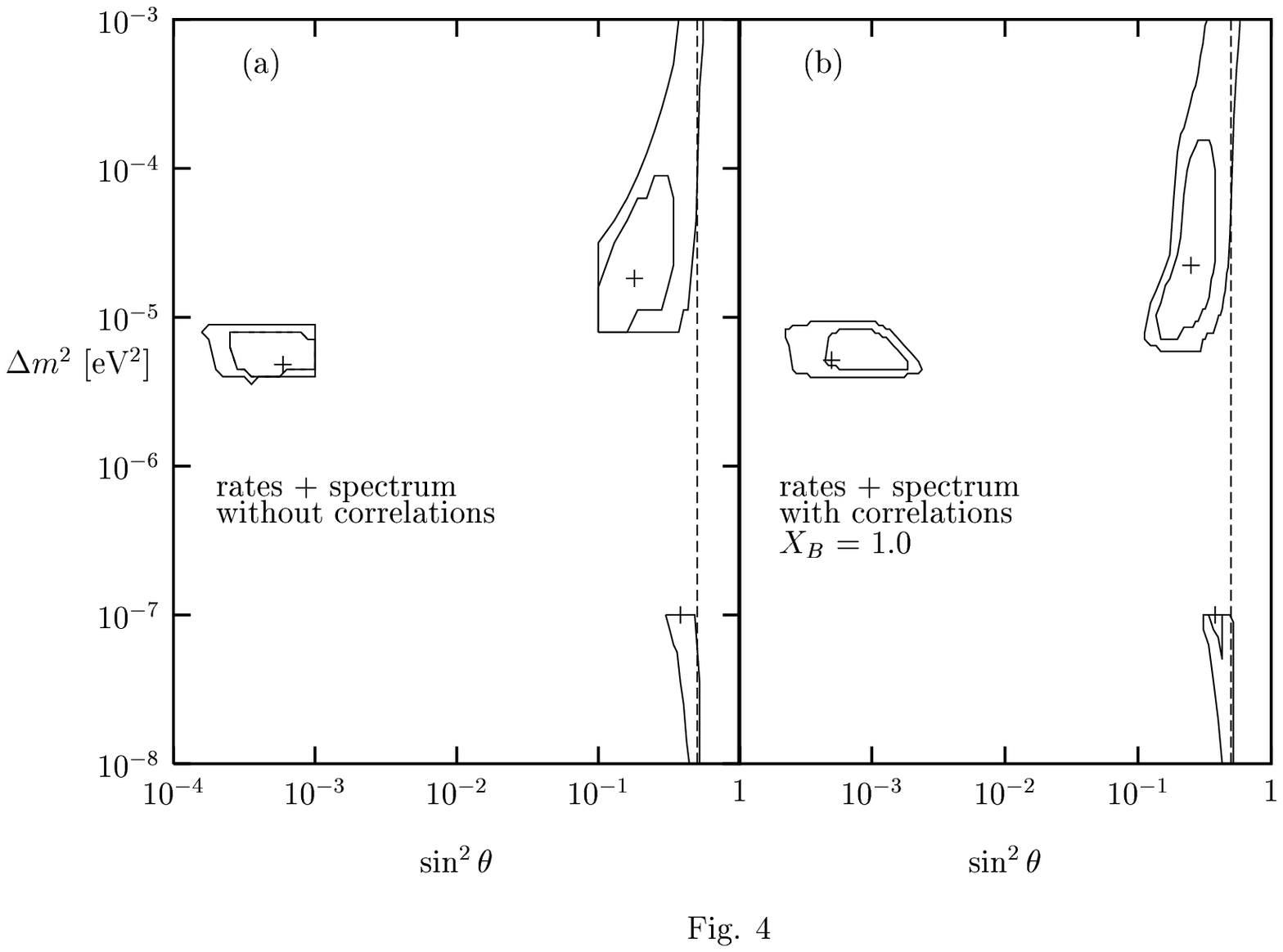}
\end{figure}

\end{document}